\documentclass[longauth]{aa} 
\usepackage{graphicx}
\usepackage{txfonts}
\begin{document}

\title{MAGIC observations and multiwavelength properties of the quasar 3C~279 in 2007 and 2009}


\author{
 J.~Aleksi\'c\inst{1} \and
 L.~A.~Antonelli\inst{2} \and
 P.~Antoranz\inst{3} \and
 M.~Backes\inst{4} \and
 J.~A.~Barrio\inst{5} \and
 D.~Bastieri\inst{6} \and
 J.~Becerra Gonz\'alez\inst{7,}\inst{8} \and
 W.~Bednarek\inst{9} \and
 A.~Berdyugin\inst{10} \and
 K.~Berger\inst{7,}\inst{8}\inst{*}
% ...
}
\institute { IFAE, Edifici Cn., Campus UAB, E-08193 Bellaterra, Spain
 \and INAF National Institute for Astrophysics, I-00136 Rome, Italy
 \and Universit\`a  di Siena, and INFN Pisa, I-53100 Siena, Italy
 \and Technische Universit\"at Dortmund, D-44221 Dortmund, Germany
 \and Universidad Complutense, E-28040 Madrid, Spain
 \and Universit\`a di Padova and INFN, I-35131 Padova, Italy
 \and Inst. de Astrof\'{\i}sica de Canarias, E-38200 La Laguna, Tenerife, Spain
 \and Depto. de Astrof\'{\i}sica, Universidad de La Laguna, E-38206 La Laguna, Spain
 \and University of \L\'od\'z, PL-90236 Lodz, Poland
 \and Tuorla Observatory, University of Turku, FI-21500 Piikki\"o, Finland
% ...
}
\date{Received 11 January 2011}

\abstract
  {3C~279, the first quasar discovered to emit VHE $\gamma$-rays by the MAGIC
  telescope in 2006,  was reobserved by MAGIC in
  January 2007 during a major optical flare and from December 2008 to
  April 2009 following an alert from the Fermi space telescope on an
  exceptionally high $\gamma$-ray state.}
  {The January 2007 observations resulted in a detection on January 16 with significance
  5.2$\sigma$, corresponding to a F($>150$\,GeV) $(3.8\pm0.8)\cdot10^{-11}$ ph cm$^{-2}$ s$^{-1}$ 
  while the overall data sample does not
  show significant signal. The December 2008 - April 2009 observations
  did not detect the source. We study the
  multiwavelength behaviour of the source at the epochs of 
  MAGIC observations, collecting quasi-simultaneous data at optical and X-ray
  frequencies and for 2009 also $\gamma$-ray data from {\it Fermi}.}
  {We study the light curves and spectral energy distribution of the 
source. The spectral energy distributions of three observing epochs 
(including the February 2006, which has been previously published) 
are modelled with one-zone inverse Compton models 
and the emission on January 16, 2007 also with two zone model and with 
a lepto-hadronic model.} 
  {We find that the  VHE $\gamma$-ray emission detected
  in 2006 and 2007 challenges standard one-zone model, based on relativistic electrons 
  in a jet scattering  broad line region photons, while the other studied models fit the observed spectral energy distribution more satisfactorily. 
  }
  {}

                            
\titlerunning{3C~279}
                             
\authorrunning{Aleksi\'c et al.}
                            
\maketitle

\section{Introduction}

3C~279 was the first quasar discovered as a $\gamma$-ray source with
the Compton Gamma-Ray Observatory (Hartman et al. 1992),
and the first
flat-spectrum radio quasar (FSRQ) discovered to
emit very high energy (VHE, defined here $>100$\,GeV) $\gamma$-rays
(Albert et al. 2008a). With a redshift of 0.536 (Hewitt \& Burbridge 1993),
3C~279 is also the most distant of the VHE $\gamma$-ray emitting
sources  discovered so far.

3C~279 is one of the brightest quasars at all wavelengths and its
multiwavelength behavior and jet structure has been studied in great
detail 
(e.g. Hartman et al. 2001, Chatterjee et al. 2008).
Its relativistic jet, which is the source of the radio to
VHE $\gamma$-ray emission, is closely aligned with the line of sight
(the angle varies, but is sometimes as small as $<0.5^\circ$,
Jorstad et al. 2004). The radio to optical emission is synchrotron
radiation emitted by the relativistic electrons spiraling in the
magnetic field of the jet. In this low-energy regime the total flux
density variations are well described by shocks propagating in the jet
(e.g. Lindfors et al. 2006). 
The X-ray emission can be explained by
the synchrotron self-Compton mechanism (SSC, e.g. 
Maraschi et al. 1992, Hartman et al. 2001, Sikora et al. 2001), where the synchrotron photons emitted by the jet act
as seed photons for the inverse Compton scattering. 

However, there is no consensus about the emission mechanism and site
of the $\gamma$-ray and VHE $\gamma$-ray emission in 3C~279. The
emission can, in principle, be explained by both leptonic and hadronic
models: the leptonic models mostly rely on external Compton (EC, e.g.
Hartman et al. 2001, Albert et al. 2008a),
invoking the inverse Compton
scattering of external photons from accretion disk or broad line
region (BLR) clouds (Dermer \& Schlickeiser 1993, Sikora et al. 1994)
, while in the hadronic
models 
(Mannheim \& Biermann 1992, M\"ucke et al. 2003, B\"ottcher et al. 2009)
the VHE $\gamma$-ray
photons are produced by proton initiated cascades or directly through
proton synchrotron radiation 
(but see 
Sikora et al. 2009 for criticisms). 

The leptonic models are very sensitive to the site of the emission: the external Compton models relying on photons
originating from broad line emission clouds are not efficient if the
emitting blob is outside the BLR 
and in the SSC models the $\gamma$-ray emission must originate from a different emission region than the main component of the synchrotron radiation in order to reproduce the observed $\gamma$-ray flux (B\"ottcher et al. 2009).
It should also be noted, that independently of the emission mechanism the
internal absorption cannot be neglected if the emission region is
located inside the BLR
(Sitarek \& Bednarek 2008, Tavecchio \& Mazin 2009). Alternatively, the emission can be produced in
regions located far beyond the broad line region, at distances at
which the dominant radiation field for EC is that of the parsec-scale
dusty torus (Sikora et al. 2008). 
In this case, internal absorption can
be neglected up to $\sim$ 1 TeV but, due to the large size of the
emission region, a minimum variability timescale of the order of
$\sim$1 day is expected.

Amongst {\it Fermi} detected blazars (Abdo et al. 2009)
hard overall spectrum (spectral index=2.32) but also the comparatively
weak evidence of a break in the spectrum at 
 1 GeV. The spectral index after the break energy is given as 2.50
  without apparent cut-off. This is one of the hardest spectrum
  of all bright FSRQs above a few GeV and makes it one of the prime target for VHE $\gamma$-ray observations.

In this paper we present VHE $\gamma$-ray observations of 3C~279
performed by the MAGIC-I telescope in January 2007 and from
December 2008 to April 2009. The 2007 observations were triggered by an
optical outburst in the source while the December 2008 observations
were triggered by an alert from the {\it Fermi} space telescope. 
The January-April 2009 data were taken as a part of the followed
multiwavelength campaign. We present the data analysis and results 
of the MAGIC observations, together with simultaneous 
multiwavelength observations and a discussion  of theoretical models. 
For comparison and completeness we also summarize the 2006 
observation campaign (Albert et al. 2008a).
The plan of the paper is as follows. The MAGIC observations, data analysis 
and  results are presented in Section 2. Multiwavelength observations
are presented in Section 3 and discussed in Section 4 together with the MAGIC data.
VHE $\gamma$-ray and multifrequency data are combined in Section 5 to build
 quasi-simultaneous spectral  energy distributions at different epochs
 and theoretical models are presented and critically discussed.
 A summary of results and conclusions are given in Section 6.

\section{MAGIC Observations}
The data described in this paper were taken with the first MAGIC 
(Major Atmospheric Gamma-ray Imaging Cherenkov) telescope as a
standalone imaging atmospheric Cherenkov telescope. It is located on the
Canary Island of La Palma. MAGIC has a standard trigger threshold of
60\,GeV for observations at low zenith angles, an angular resolution
of $ \sim 0.1^\circ$ on the event by event basis and an energy
resolution above 150\,GeV of $\sim 25\%$ 
(see Albert et al. 2008b for details).

MAGIC observed 3C~279 in January 2007 and from December 2008 to
April 2009. Due to changes in the telescope systems these data sets were
analyzed separately. 

\subsection{January 2007}

3C~279 was observed during nine nights in January 2007 for a total of
23.6\,h, 18.6\,h (seven nights) of which passed the quality selection. The
observations were conducted in moon, dark night and twilight. The zenith angle range of the observations was 35 to 46
degrees. The readout chain included a fast 300\,MSample/s FADC system,
which allows us to use the time information of the showers in the
analysis as described in Aliu et al. (2009).
Since the average night sky background noise level stayed below 3 
times the dark time level, all data has been analyzed with the same 
standard image cleaning levels as discussed in 
Britzger et al. (2009). The data were taken in On mode where
the telescope is pointing directly towards the source. So-called
Off data were taken under similar conditions (zenith angle, trigger 
rates and weather conditions) from September 2006 until  January 2007 and has 
been used to estimate the background level. Due to the installation of 
the new 2\,GHz MUX-FADC read-out system in 2007 February, no Off data 
after January 2007 were fulfilling the requirement of similar 
observation and hardware conditions.

The 2007 data were analyzed using so-called W\"urzburg Analysis chain
described in detail in Bretz \& Wagner (2003) and Bretz \& Dorner
(2008).
The data were calibrated following the
description in 
Albert et al. (2008c), the signal was extracted at the
pulse maximum using a spline method, the air-shower images were
cleaned of noise from night-sky background light by applying a
time image cleaning. In the first
step a minimum number of 6 photo-electrons in two adjacent pixels is required
(so called "core pixels"). Each pixel next to a core pixel, which is above a
threshold of 3 photo-electrons is considered a "boundary pixel". Additionally
to these limits in the charge of the pixels it is required that the arrival
time of core pixels is within 1.75\,ns of the mean arrival time of all core
pixels and that boundary pixels arrive within 4.5\,ns of the mean arrival
time of the core pixels. Both the charge and the time limits have to be
fulfilled, otherwise the pixel is considered to contain noise and deleted
from the image. We parametrized the shower images 
(Hillas, 1985) 
and used a SIZE-dependent parabolic cut in AREA, WIDTH and LENGTH 
(Riegel et al. 2005)  
for the $\gamma$/hadron separation. The cut parameters were
optimized on Crab Nebula data taken within the same
zenith angle range as the 3C~279 data. The energy threshold of the 
analysis was 220\, GeV. 
The arrival direction
of the $\gamma$-rays is reconstructed using the DISP method 
(Fomin et al., 1994; Lessard et al., 2001), 
which was adapted to use the shower timing information as described in 
Aleksi\'c et al. (2010).  
The significance of a detection is evaluated by comparison of the number
of events in the On and the Off data sample using formula 17 of 
Li \& Ma (1983).

\begin{table*}[!ht]
\begin{center}
\caption{Night-by-night results of MAGIC January 2007 observations of 3C~279.}
\scriptsize
\begin{tabular}{| l | l | l | l | l | l |}
  \hline
  Observation night [MJD] & Observation time [min] & Excess events [counts]& background events [counts]& On/Off scaling factor$^{1}$ & significance$^{2}$ \\ \hline
  54115 & 149.5 & 17.6 & 97.4 & 0.058 & 1.7 \\ \hline
  54116 & 151.1 & 64.1 & 102.9 & 0.062 & 5.6 \\ \hline
  54117 & 157.2 & -7.9 & 100.9 & 0.060 & -0.8 \\ \hline
  54118 & 153.6 & -6.0 & 82.0 & 0.049 & -0.7 \\ \hline
  54120 & 164.1 & -12.3 & 117.3 & 0.07 & -1.1 \\ \hline
  54121 & 166.7 & 12.7 & 119.3 & 0.07 & 1.1 \\ \hline
  54122 & 175.1 & 18.2 & 126.8 & 0.076 & 1.5 \\ \hline
 \hline
\end{tabular}
\end{center} 
\begin{tabular}{l}
\tiny
$^1$The scaling factor is the ratio between the on and the off $\theta^²$ distribution normalized outside the signal region.\\
$^2$Significance is given as standard deviations and is not corrected for seven trials, arising from diving the data sample by seven \\(corresponding to seven nights with good quality data).
\end{tabular}
\end{table*}
\normalsize

Since in 2006 the source was detected in a single day flare, every
observation day was analyzed separately. Table 1 summarizes the
analysis results of the January 2007 campaign. Out of the seven
observation nights with good data quality only one, MJD 54116
(January 16), revealed a significant excess in the MAGIC data after
Off-subtraction. The significance of the excess is 5.6$\sigma$
(pre-trial) with 64.1 excess events above 102.9 normalized Off events
in 151\,min of observations. The $\theta^2$ (squared distance between
the true and reconstructed source position, see e.g. Daum et al. 1997)
distribution is shown in Fig.~1. None of the other observation nights
yielded in a significant excess.

\begin{figure}
\includegraphics[width=0.45\textwidth]{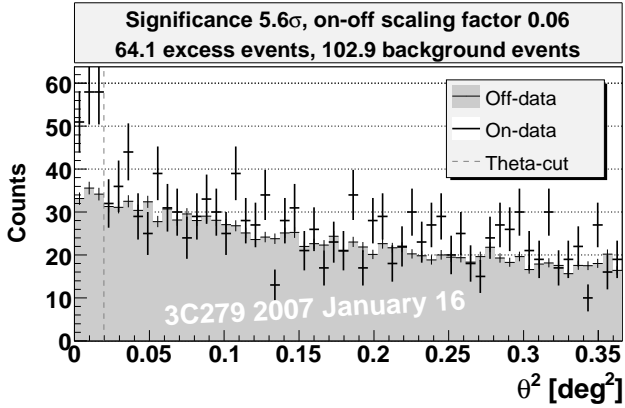}
\caption{Reconstructed shower direction $\theta^2$ for the
Off (gray shaded area) and On (black crosses) data
as observed on January 16, 2007 by MAGIC.
The vertical dotted line corresponds to the apriori defined signal
region. An excess of 64 events is clearly visible in the
On data as compared to the normalized Off data. The corresponding 
pre-trial significance is 5.6$\sigma$
and 5.4$\sigma$ after correction for seven trials, respectively.}
\label{theta_january07}
\end{figure}

\begin{figure}
\includegraphics[width=0.45\textwidth]{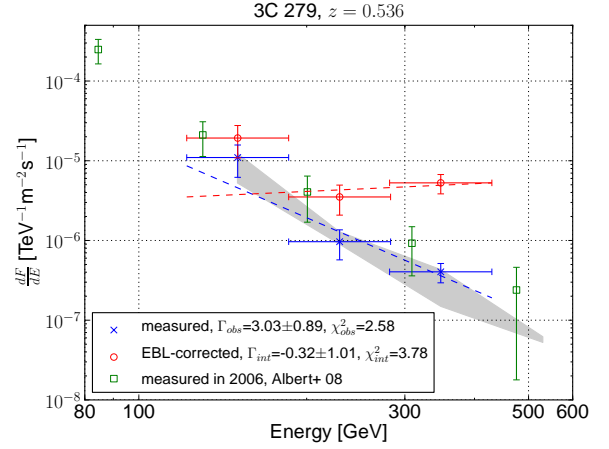}
\caption{Spectrum of the VHE $\gamma$-ray flare January 16, 2007. 
The {\bf blue} points correspond to the 
measured spectrum with the respective errors on the flux and the
bin size in energy. The results of a simple power law fit are shown 
as blue line.
The gray shaded region represents the systematic 
error of the analysis. The red points show the spectrum deabsorbed using the EBL model by Dominguez et al. (2010) and the red dashed line the fit to this EBL corrected spectrum. For comparison the spectrum from 2006 (Albert et al. 2008a) is shown with green points.}
\label{spectrum_2007}
\end{figure}

For spectrum (Figure~2) and light curve softer cuts that have a higher
$\gamma$-ray efficiency were used. 
The VHE $\gamma$-ray spectrum of the flare can
be described by a simple power law (with the differential flux
given in units of TeV$^{-1}$ m$^{-2}$ s$^{-1}$):
\begin{equation}
\frac{\mathrm {d}F}{\mathrm{d}E} = (5.7\pm 1.3)\times10^{-7}\big(\frac{\mathrm E}{300\,\mathrm {GeV}}\big)^{-3.1\pm1.1} 
\end{equation}
The highest spectral energy point has a medium energy of
350\,GeV. For a Crab like spectrum we have a systematic energy scale 
error of
16\%, a systematic error of 11\% on the flux normalization
(without the energy scale error), and a systematic slope error
of 0.2 (a detailed list of all the contributions can be found in 
Albert et al. 2008b), while for the soft spectrum like 3C~279 the systematic errors are expected to be slightly larger. 
Assuming that 3C~279 always emits $\gamma$-rays above 150 GeV we calculated
the corresponding light curve (see Section 4) on the night by night basis. 
For January 16th the integral flux above 150~GeV is $(3.8\pm0.8)\cdot10^{-10}$ ph cm$^{-2}$s$^{-1}$.

The results were cross-checked using the standard MAGIC analysis chain 
(Albert et al. 2008c). The standard analysis resulted in lower significance, 4.6$\sigma$, for the detection. For flux and spectral index the results were comparable for both analysis chains.

\subsection{December 2008- April 2009}
\begin{table*}[!ht]
\begin{center}
\caption{Night-by-night results of MAGIC December 2008 -- April 2009 observations.}
\scriptsize
\begin{tabular}{| l | l | l | l | l | l |}
   \hline
   Observation night [MJD] & Observation time [min] & Excess events [counts]& 
background events [counts]& On/Off scaling factor$^{1}$& significance$^{2}$ \\ \hline
   54829 & 50.4 & -0.3  & 65.3& 0.03 &-0.04  \\ \hline
   54832 &60.0 & -10.7 &  92.7 & 0.02 & -0.98 \\ \hline
   54834 & 81.0 & 4.8 & 103.2&  0.02 & 0.41 \\ \hline
   54838 & 78.0 &-1.4  & 48.4&  0.08 & -0.18 \\ \hline
   54852 & 33.6 &12.9  & 42.1& 0.07 &1.73  \\ \hline
   54855 &157.8  & 4.1 & 241.9 & 0.06 & 0.24 \\ \hline
   54856 &140.4 &5.3  &70.7 & 0.10  &0.57 \\ \hline
   54937 &114.6 & -0.3 &110.3&  0.03 &-0.03  \\ \hline
   \hline
\end{tabular}
\end{center}
\begin{tabular}{l}
\tiny
$^1$The scaling factor is the ratio between the on and the off $\theta^²$ distribution normalized outside the signal region.\\
$^2$Significance is given as standard deviations and is not corrected for trials, arising from diving the data sample by eight \\(corresponding to eight nights with good quality data).
\end{tabular}
\end{table*}
\normalsize

MAGIC observations started on December 9, 2008 after the {\it Fermi}
Collaboration announced a high GeV $\gamma$-ray state of the source
(Ciprini et al. 2008).
Unfortunately 3C~279 could only be observed at zenith angles larger than
46 degrees and for a short time ($\sim$10\,min) due to
visibility constraints.  These short runs were insufficient to detect the source. Follow-up observations were conducted at the end of 
December 2008 until April 2009 under more favorable zenith angles, smaller than
35 degrees. 

A total observation time of 28.1\,h was accumulated over 20 days with
the 2\,GHz MUX-FADC read-out system.  35\,min were recorded in
On-mode, while the remaining data (27.5\,h) were taken in wobble mode
where the source was displaced by 0.4\,$^\circ$ from the camera center
in order to allow the recording of simultaneous Off data with the same
offset from the camera center (Daum et al. 1997). 
After quality selection
11.9\,h of data were used in the analysis. The main part of this data
set is from January 2009, but it also includes one night (29) in
December 2008 and one (16) in April 2009.

The data were analyzed using the standard analysis chain as described in
Albert et al. (2008b, 2008c) and Aliu et al. (2009).
In order to suppress the background showers produced by charged
cosmic rays, a multivariate classification
method known as Random Forest is used (Albert et al. 2008d). 
For every
event, the algorithm takes as input a set of image
parameters, and produces one single parameter as output,
called \textsc{Hadronness}.
The background rejection is then achieved by a cut in \textsc{Hadronness},
which was optimized using Crab Nebula data taken
under comparable conditions.

A cut in $\theta^2$, was used to extract the signal and
also optimized in the same way. An additional cut removed the events
with a total charge of less than 200 photo-electrons (phe) in order to
assure a better background rejection.  

We find no signal in this data set. The number of excess events is 29
over a background of 775 events, corresponding to a significance 
of 0.94$\sigma$. The night by night results are reported in Table~2.

\begin{table}[!h]
\begin{center}
\caption{Differential upper limits (95\% confidence level) for December 2008-April 2009 observations.}
\begin{tabular}{| l | l |}
   \hline
Energy bin[GeV] &           Upper Limit[1/TeV/m2/s] \\ \hline 
150-200         &             1.8$\cdot10^{-5}$          \\ \hline
200-315         &             1.6$\cdot10^{-6}$          \\ \hline
315-430         &             7.2$\cdot10^{-7}$         \\ \hline
430-580         &             2.5$\cdot10^{-7}$         \\ \hline
\end{tabular}
\end{center}
\end{table}

All data were combined for the calculation of  flux upper limits in
the energy range between 150\,GeV and 580\,GeV (higher energies can be
neglected due to the strong EBL absorption).  Including a 30$\%$
systematic error and assuming a power law with a spectral index of -4
the integrated flux upper limit (2$\sigma$ c.l.) is 6.3$\cdot 10^{-7}$
ph/m$^2$/TeV/s. In Table 3 the differential values are reported.

\section{Multiwavelength Observations}
We present here long term optical monitoring data from the Tuorla
blazar monitoring program, and near infrared data from REM (for
January 2007), X-ray data from SWIFT (for February 1, 2009) and $\gamma$-ray
data from {\it Fermi} (for January 2009).

\subsection{Optical Observations}
3C~279 has been observed regularly since 2004 as a part of the Tuorla
blazar monitoring program.\footnote{http://users.utu.fi/kani/1m/} Figure 3 shows the long term optical R-band
light curve covering all the MAGIC observation periods discussed in
this paper. In 2009 also polarimetric observations of 3C~279 were
performed. The photometric optical R-band observations are performed
with the Tuorla 1 meter telescope (Finland) and Kungliga
Vetenskapsakademien (KVA) 35 cm telescope (La Palma). The latter
can be controlled remotely from Tuorla Observatory.

\begin{figure}[!ht]
\includegraphics[angle=270, width=0.45\textwidth]{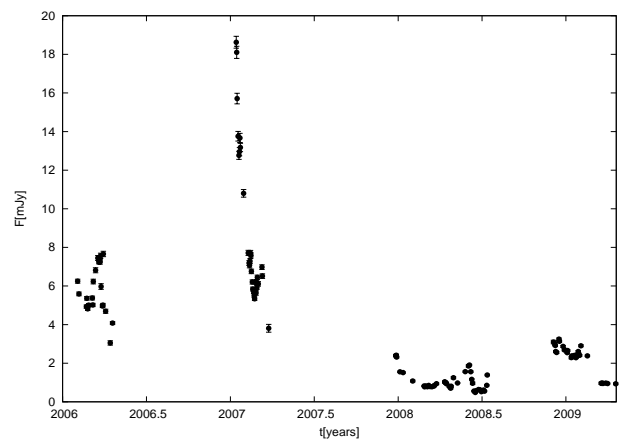}
\caption{Long term optical R-band observations of the Tuorla blazar monitoring program of 3C~279. The giant flare near the beginning of
2007 is clearly visible.}
\label{LC_opt_all}
\end{figure}

KVA consists of two telescopes, the larger one being a 60cm (f/15)
Cassegrain telescope equipped with a CCD polarimeter capable of
polarimetric measurements in BVRI-bands using a plane-parallel calcite
plate and a super-achromatic lambda/2 retarder.  For 3C~279 the
observations were done without filter to improve the signal-to-noise
of the observations.

The 35cm telescope is used for photometric measurements in B, V and R-bands.
The photometric measurements are made in differential mode, i.e. by
obtaining CCD images of the target and calibrated comparison stars in
the same field of view (Raiteri et al. 1998). 
The magnitudes of the source and comparison stars are measured using
aperture photometry and the (color corrected) zero point of the image
determined from the comparison star magnitudes. Finally, the object
magnitude is computed using the zero point and a filter-dependent
color correction. Magnitudes are then transferred to linear fluxes
using the formula $F = F_0 \cdot 10^{\mathrm{mag}/-2.5} $, where mag is the object
magnitude and $F_0$ is a filter-dependent zero point (in the R-band the
value $F_0 = 3080$ Jy is used from Bessell et al. 1979).

For polarimetric measurements the normalized Stokes parameters and the
degree of polarization and position angle were calculated from the
intensity ratios of the ordinary and extraordinary beams using
standard formula 
(e.g. Degl'Innocenti et al. 2007) 
and semiautomatic
software specially developed for polarization monitoring
purposes. During some of the nights, polarized standard stars from
Turnshek et al. (1990) 
were observed to determine the zero point of the
position angle. The instrumental polarization of the telescope has
been found to be negligible.

\subsection{REM observations in January 2007}

In January 2007 3C~279 was intensively monitored by REM in the IR band (J, H
and K filters). REM is a 60-cm diameter fast-reacting telescope with
10$^\circ$\,s$^{-1}$ pointing speed located at the Cerro La Silla
premises of the European Southern Observatory, Chile
(Zerbi et al. 2001;
Chincarini et al. 2003; Covino et al. 2004a; Covino et al. 2004b).
The telescope hosts REMIR, an
infrared imaging camera, and ROSS, an optical imager and slitless
spectrograph. The two cameras observe simultaneously the same field of
view of $10$\arcmin$\times10$\arcmin\ thanks to a dichroic mirror, although
only the infrared data are reported here. The telescope was designed
to rapidly follow transient high-energy events as Gamma-Ray Bursts
(GRBs) but in the idle time it is perfectly suited for monitoring
programs of variable sources. The observations, including the short
and long-term scheduling, are totally automatic.  Typical exposure
durations were of 150\,s in the $J$, $H$ and $Ks$ filters. Data were
reduced in a standard way by means of tools provided by the
ESO-Eclipse package (Devillard 1997).
Standard aperture photometry was
derived and results calibrated by a suitable number of well-exposed
2MASS objects in the
field.\footnote{http://www.ipac.caltech.edu/2mass/}

\subsection{Swift and Fermi observations in January-February 2009}
The {\it Swift} satellite (Gehrels et al. 2004) 
is a NASA mission, launched in 2003. 
{\it Swift} performed Target of Opportunity 
observation of 3C~279 on February 1, 2009.

{\it Swift}/XRT is a Wolter type I grazing incidence telescope, with
110 cm$^2$ effective area, 23.6' arcminutes FOV and 15'' angular
resolution, sensitive in the 0.2--10 keV energy band.  Data were
reduced using the software distributed with the {\tt heasoft} 6.3.2
package by the NASA High Energy Astrophysics Archive Research Center
(HEASARC). The {\tt xrtpipeline} was set for the photon counting or
window timing modes and having selected single pixel events (grade
$0$).  Data shown in the SEDs were rebinned in order to have at least
30 counts per energy bin.  Power law models have been fitted to the
spectra. The X-ray reddening due to absorbing systems along the light
travel path has been corrected assuming the Galactic value for column
density of neutral hydrogen $N_{H} = 2.1\times 10^{20}$ cm$^{-2}$
(Kalberla et al. 2005).

UVOT is a 30 cm diffraction limited optical--UV telescope, equipped
with 6 different filters, sensitive in the 1700--6500 \AA\ wavelenght
range, in a 17' $\times$ 17' FOV. During the pointing of February
1, 2009 used below for the SED, only filters UVW1 and U were available.
Analysis was performed by means of the \texttt{uvotimsum} and
\texttt{uvotsource} tasks with a source region of $5''$, while the
background was extracted from a source--free circular region with
radius equal to $50''$ (it was not possible to use an annular region,
because of a nearby source).  The extracted $\nu F_{\nu }$ magnitudes
have been corrected for Galactic extinction using the values of Schiegel et al. (1998) and applying the formulae by Pei (1992) for the UV filters,
and finally converted into fluxes following Poole et al. (2008).

The {\it Fermi} LAT is a pair conversion telescope designed to cover the
energy band from 20 MeV to greater than 300 GeV which operates in
all-sky scanning mode.

The average LAT spectrum was derived in the period of the MAGIC
pointings using the publicly available LAT data.\footnote{accessible
  from http://fermi.gsfc.nasa.gov} The photons of class 3
(DIFFUSE) with energy in the range 0.1--100 GeV collected from
January 21, 2009 to January 31, 2009 were selected. These data were processed by using
Science Tools 9.15.2, which includes the Galactic diffuse and
isotropic background and the Instrument Response Function IRF
P6\_V3\_DIFFUSE. Then photons in the good--time intervals and
within a region of interest (ROI) with radius of 10$^{\circ}$ from the
source radio position were selected and a cut on the zenith angle parameter
($<105^{\circ}$) to avoid the Earth albedo was applied. The following steps were
to calculate the live--time cube, the exposure map and the diffuse
response.
 
With all these information at hands, an analysis by using
an unbinned likelihood algorithm ({\tt gtlike}) in separate energy
bins was performed. The model included the isotropic and Galactic diffuse
backgrounds, the source of interest, all the 1FGL sources in the ROI
and, possibly, additional sources not included there but identified in
the map. For all the point sources we assumed a power law spectrum,
with flux and photon index as a free parameter and calculated the
corresponding test statistic ($TS$, see Mattox et al. (1996) for a
definition; in practice one assumes $\sqrt{TS}\simeq \sigma$, the
significance of the detection). In the SED we report the four bins
with high significance, $TS>25$.

\section{Multiwavelength behavior}

In addition to multiwavelength data discussed in the previous section,
there are also previously published multiwavelength data for all the
MAGIC observing epochs (spring 2006: B\"ottcher et al. 2007, Collmar et al. 2010, winter 2006-2007:
Larionov et al. 2008, 1996-2007: Chatterjee et al. 2008, winter
2008-spring 2009: Abdo et al. 2010). In the following we discuss the
multiwavelength behaviour of the source in these three epochs
comparing the data described in this paper to previously published
results.

In the spring of 2006 when 3C~279 was first discovered by MAGIC, the
source was in a high optical and X-ray state. However no optical or
X-ray flare simultaneous to the VHE $\gamma$-ray flare was observed (Fig.~4).
We are not aware of any optical polarization monitoring data for this
epoch.  In the radio bands the source was in quiescent state
(B\"ottcher et al. 2007, Collmar et al. 2010) and there was no coincident very long baseline interferometry (VLBI)
knot emerging from the VLBI core (Chatterjee et el. 2008), suggesting
that the VHE $\gamma$-ray emission originates from region that is
opaque to radio frequencies.

\begin{figure}
\includegraphics[width=0.45\textwidth, angle=270]{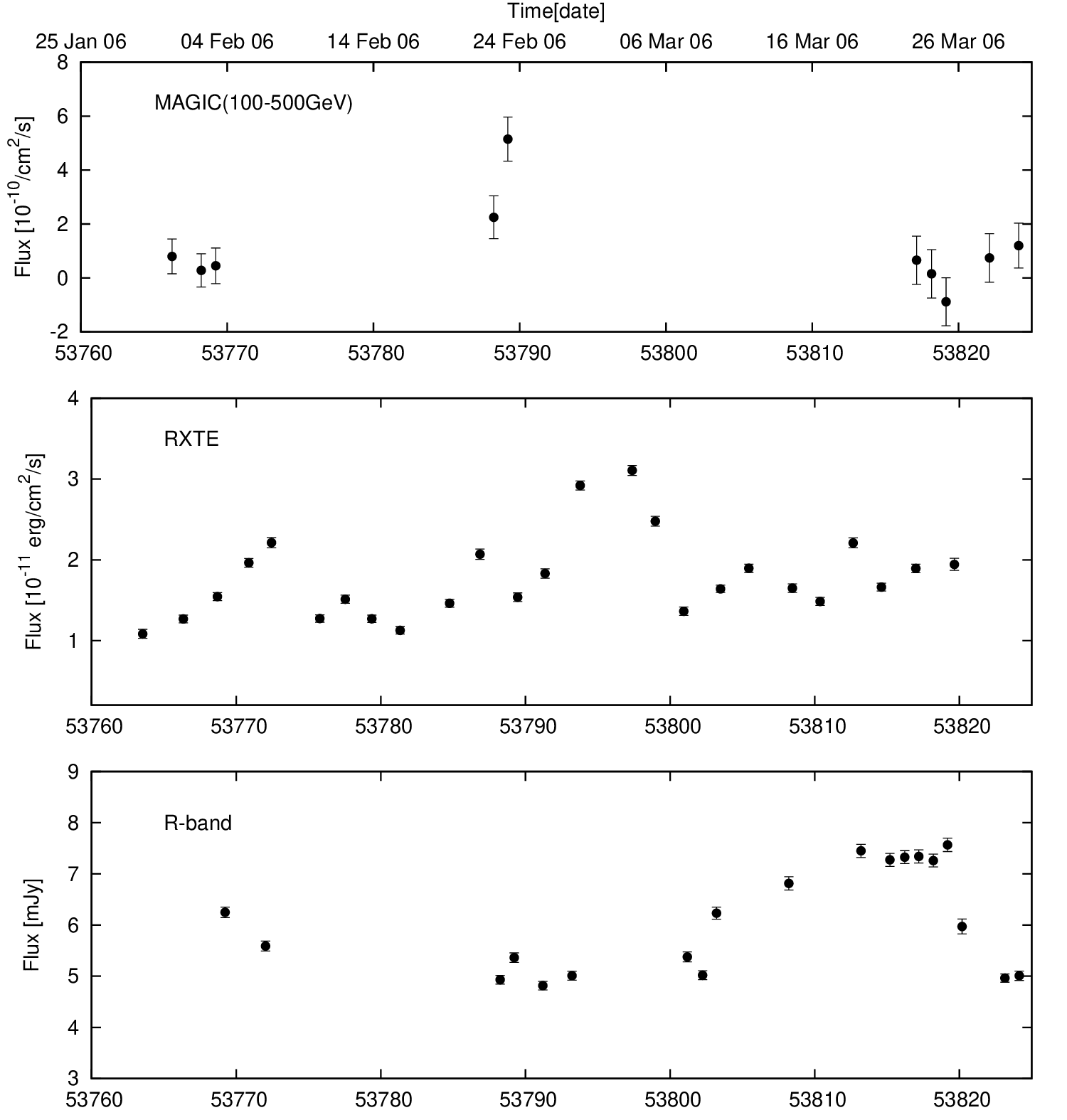}
\caption{Light curve in different energy bands during the 2006 observation campaign. From top to bottom: 
VHE $\gamma$-ray flux between 100-500 GeV as measured by MAGIC (Albert et al. 2008a), RXTE PCA flux 
in the 2-10 keV range (Chatterjee et al. 2008)
and KVA R-band
observations.}
\label{LC_2006}
\end{figure}

In January 2007 3C~279 was very bright in the optical band, reaching a
peak flux of 19 mJy (R$\sim$13) on MJD 54114 . This is the maximum
flux observed from the source in five years of Tuorla blazar
monitoring. However, during the MAGIC observation the optical flux was
already decreasing and on the night of January 16, 2007 (MJD 54116) no
increase in the optical flux was detected. A similar behavior was seen
in the near-IR band, with a clear peak at January 13
followed by a decay.  Unfortunately REM did not observe 3C~279 on
January 15 and 16, but both on January 14 and 17 the data show
a steep spectrum, smoothly joining with the R-band measure of
KVA. Therefore, it is reasonable to assume that also the IR flux
decreased smoothly during the VHE $\gamma$-ray flare.

The fluxes inferred from the MAGIC ($>$150 GeV), the RXTE PCA
(2--10~keV), the KVA (R-band) and REM near infrared observations of
January 2007 are compared as a function of time (light curves) in
Fig.~5.  As can be seen in Fig.~5, the VHE $\gamma$-ray flare does not
coincide with neither the maximum of the optical flux or the maximum
of the X-ray flux that were both observed prior to MAGIC
observations. A comparison of the different light curves suggests a
possible lag of 2-3 days of the VHE $\gamma$-ray flare with respect to
the lower frequencies. It should also be noted, that the X-ray flux is
in general lower than during the MAGIC observations in 2006.

\begin{figure}
\includegraphics[width=0.45\textwidth]{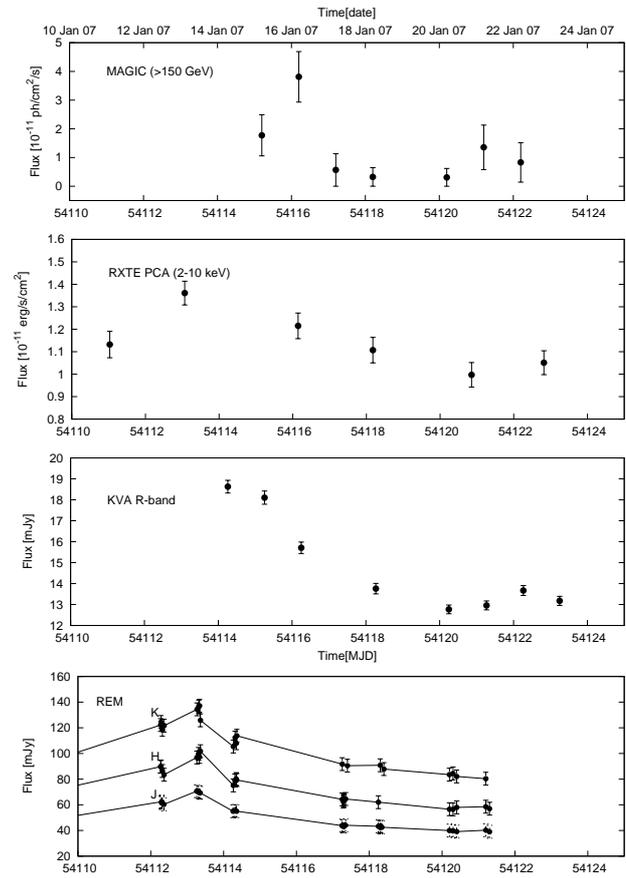}
\caption{Light curve in different energy bands during the 2007 observation campaign. From top to bottom: 
VHE $\gamma$-ray flux above 150 GeV as measured by MAGIC, RXTE PCA flux 
in the 2-10 keV range (Chatterjee et al. 2008),
KVA R-band
observations and REM infrared observations. For the MAGIC light curve the night by night flux is calculated assuming that 3C~279 always emits $\gamma$-rays above 150~GeV.}
\label{LC_2007}
\end{figure}

In 2006-2007 there was also extensive whole earth blazar telescope
(WEBT), very long baseline array (VLBA) and RXTE X-ray satellite
monitoring of 3C~279 (Larionov et al. 2008). The VLBA data shows a
component emerging from the core at MJD $54063\pm40$ accompanied by
the maxima in the 37\,GHz light curve while the optical polarization data
shows a rotation of the polarization angle $\sim300^\circ$ starting
around MJD 54115 with duration of $\sim2$ months. There is coincident
rotation of the polarization angle of the VLBI 43 GHz core and thus
Larionov et al. (2008) conclude that the optical flares and rotation
of the optical polarization angle take place in the VLBI knot as it
moves downstream of the core. The timing of the MAGIC detection (MJD
54116) suggests that the emission of the VHE $\gamma$-rays also
happens in this same emission region. This would be in agreement with
previous results, that the $\gamma$-ray flares observed by EGRET and
{\it Fermi} would take place in the knots freshly emerging from the
VLBI core 
(see Jorstad et al. 2001,
Jorstad et al. 2010, 
and Agudo et al. 2011). This is sometimes denoted as 'far dissipation'
scenario, as the VLBI core is located tens of parsecs from the central
engine (e.g. Marscher et al. 2010a). However, it has been argued
(e.g. Tavecchio et al. 2010) that the size of the emission region at
this distance from the central engine is rather large which might be
in contradiction with the day scale variability of the MAGIC
observation.
This is further discussed in the following sections.

\begin{figure}
\includegraphics[width=0.45\textwidth, angle=270]{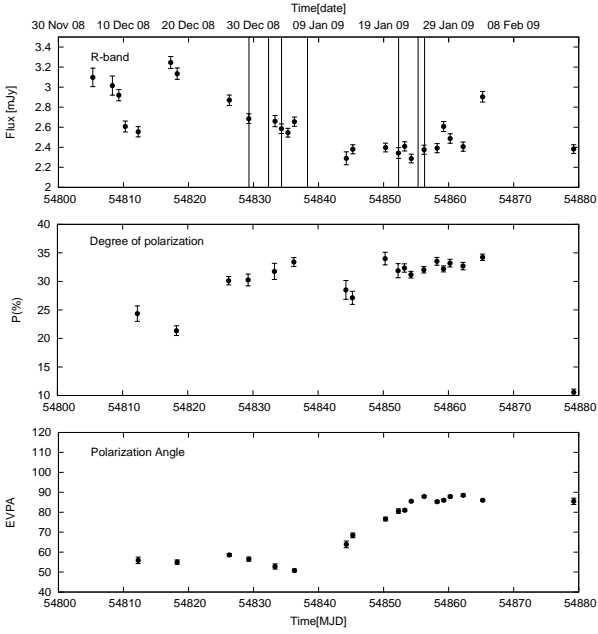}
\caption{The optical R-band light curve (top), degree of polarization (middle) and polarization angle of 3C~279 from 2008 December to 2009 February. The vertical lines show the time of MAGIC observations.}
\label{LC_2009}
\end{figure}

In December 2008 - January 2009 the source was in a quiescent state in
the optical band. The polarization degree remained rather high
($\sim$30\%) during the MAGIC observations and the polarization angle
was rotating smoothly from 60 to 90 degrees. The light curves are shown
in Figure~6. In the X-ray band the source was at quiescent level, the
flux being about 50\% lower than in January 2007 (Abdo et al.
2010). In the $\gamma$-ray band ({\it Fermi}) the source was active
from mid-November 2008 to the end of February 2009 and showed two
major flares (first one peaking end of November (MJD 54795) and second
mid February (MJD 54880)), but the core of the MAGIC observations in 2009
January (MJD 54829-54856) took place in between the two flares, when
the $\gamma$-ray flux was at a relatively low level.  Unfortunately,
due to bad weather on the MAGIC site, no data could be taken in
February, when {\it Fermi} observed the $\gamma$-ray flare which was
accompanied by a rotation of the optical polarization angle nor at the
end of April when the source was flaring in the X-ray band (Abdo et
al. 2010). 

The optical to VHE $\gamma$-ray data of the three MAGIC observing
epochs is summarized in Fig.~7. In summary, although the detections
at VHE $\gamma$-ray energies correspond to high states in optical and
X-rays, no multiwavelength correlations on short timescales seem to be
present.

\begin{figure}
\includegraphics[width=0.45\textwidth]{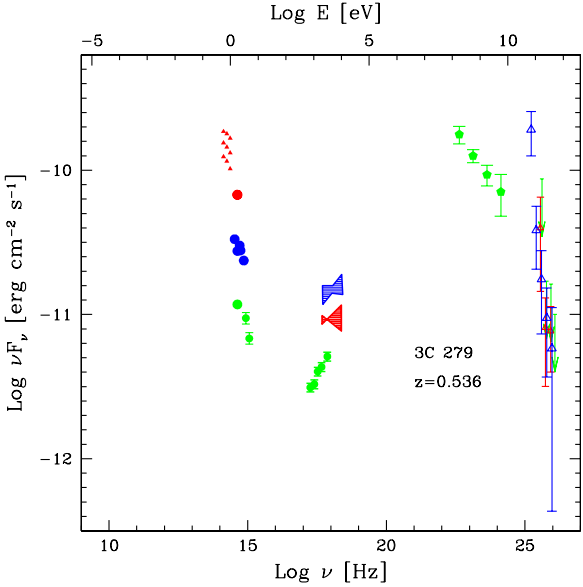}
\caption{SEDs of simultaneous optical, X-ray and $\gamma$--ray data at 
the epochs of MAGIC observations. Blue: February 23, 2006, red: January 16, 2007, green January 21-February 1 2009.}
\label{comparison}
\end{figure}

\section{Spectral Energy Distributions and Models}

In the following we discuss the spectral energy distributions (SEDs)
of 3C~279 at three epochs, February 23, 2006, January 16, 2007 and
January 21-February 1, 2009. The first epoch corresponds to the MAGIC
discovery of the VHE $\gamma$-ray emission from 3C~279 reported in
Albert et al. (2008a). 
In addition to the multiwavelength data discussed there we show IR,
optical and X-ray data simultaneous to the MAGIC observations from
B\"ottcher et al. (2009). 
For January 16, 2007 and January 21-31, 2009 we show the
quasi-simultaneous data described in the Section 3.

The optical-IR data are important for SED modeling, since the steep
continuum revealed by these measurements constrain the position of the
synchrotron peak below the optical band in 2006 and below the IR band
in 2007.  The X-ray spectral data are also quite constraining, since
they allow us to define the low energy end of the inverse Compton
component.

The spectrum observed by MAGIC 
was corrected for intergalactic absorption using the recent EBL model by
Dominguez et al. (2010) adopted here throughout. In the original publication of the 2006
observations the EBL model by Primack et al. (2005) 
was used, which underestimates
the IR background according to observations (e.g. Fazio et al. 2004; Chary et al. 2004). However, as discussed in Dominguez et al. (2010), the difference between the EBL models is smaller than the systematic uncertainties of the MAGIC data analysis.

\subsection{ SED February 2006}
 
We try to reproduce the revised data with a simple one-zone leptonic emission 
model for FSRQs (details in  Maraschi \& Tavecchio 2003)
considering the synchrotron and inverse Compton (IC) 
emission from a population of relativistic electrons in a single 
emission region (spherical, with radius 
$R$) in motion with bulk Lorentz factor $\Gamma$ at an angle $\theta$ with the line of sight. 
The electron energy distribution is described by a smoothed broken-power law with normalization 
$K$ (measuring the number density of electrons at $\gamma =1$) extending from $\gamma _1$
to $\gamma _2$, with indices $n_1$ and $n_2$ below and above the break at a $\gamma _b$, respectively. 
The magnetic field with intensity $B$ is supposed to be homogeneous
 and tangled. The seed photons for the IC emission are both the synchrotron photons produced within the jet
(SSC mechanism) and those outside the jet. We consider two cases, in which the IC process occurs within or outside the BLR. In the first case the high energy emission is dominated 
by comptonization of the UV photons of the BLR (EC/BLR). In the latter case the external radiation field is dominated by the IR thermal emission from the (putative) dusty torus (EC/IR, 
Ghisellini \& Tavecchio 2008, Sikora et al. 2009). In the case of the BLR emission we assume that the clouds are located at a distance 
$R_{\rm BLR}$ from the central black hole. The resulting emission is modeled as a black body 
peaking at $10^{15}$ Hz 
(Tavecchio \& Ghisellini 2008)
with  luminosity fixed to a fraction 
$\tau _{\rm BLR}$ of the disk luminosity 
(fixed to $L_D=2\times 10^{45}$ erg s$^{-1}$, Pian et al. 1999
). In the 
case of the torus  emission, located at a distance
 $R_{\rm IR}$ we assume that a fraction $\tau_{\rm IR}=0.5$ of the disk luminosity  is intercepted 
 and re-radiated from  dust as IR emission (again, we assume a black body spectrum, with 
 temperature $T_{\rm IR}=10^3$ K, see 
Nenkova et al. 2008). 

In Figure~8 the blue line indicates the SED obtained applying the first 
model (EC/BLR), while the red line corresponds to the EC/IR case. The model parameters are listed in Table 4.

\begin{figure}
\includegraphics[width=0.45\textwidth]{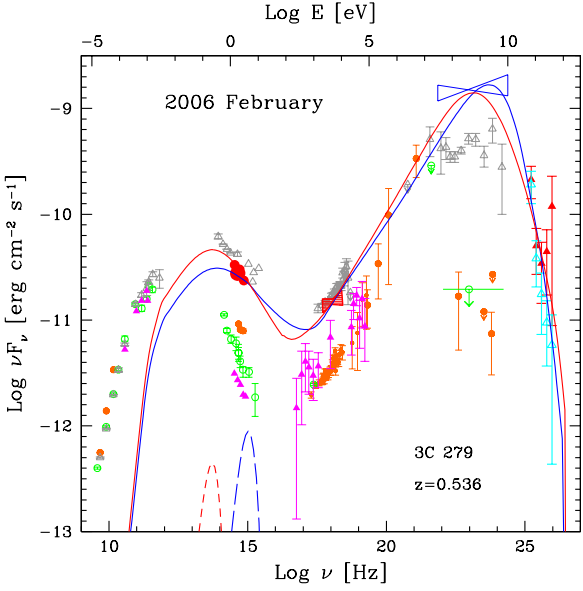}
\caption{The spectral energy distribution of 3C~279 on February 23, 2006. 
The red symbols show the optical (filled circles), RXTE (bow-tie) 
B\"oettcher (2009) and MAGIC 
deabsorbed data (triangles) 
used to fit the two model curves:
 blue line assuming EC inside the BLR and red line EC outside the BLR. 
 Also MAGIC observed data is shown (cyan). 
For comparison also historical data are shown: 1991 high state   
(gray: Hartman et al. 1996)
, 1993 low state (green: Maraschi et al. 1993)
1996 high state (orange, blue bow-tie: Ballo et al. 2002, Wehrle et al. 1998)  
and 2003 low state (magenta: Collmar et al. 2004).
The dashed lines correspond to blackbody radiation from the IR torus (red) and BLR (blue).}
\label{sed_2006}
\end{figure}

The main difference between the two models (BLR and IR) is related to
the different frequency of the target photons, determining the energy
threshold above which the IC process is strongly suppressed due to the
onset of the Klein-Nishina (hereafter KN) regime. Basically, this
energy can be written as (Tavecchio \& Ghisellini 2008, Ghisellini \&
Tavecchio 2009):

\begin{equation}
E_{\rm KN }\, = \,\frac{22.5}{\nu _{o,15}} \frac{\delta}{\Gamma (1+z)} \,\, {\rm GeV}
\end{equation}
\noindent
where $\delta$ is the relativistic Doppler factor ($\delta=\Gamma$ for $\theta=1/\Gamma$) and $\nu _{o, 15}$ 
is the frequency of the target photons (in units of $10^{15}$ Hz). Therefore, if the IC process takes place within the BLR 
the emission in the MAGIC band is strongly affected.  The  KN effect 
is more moderate if the IC scattering takes place outside the BLR. In this case the dominant photon frequency is $\nu_o=10^{13}$ Hz
 corresponding to $T_{IR}=10^3$ K; hence $E_{\rm KN }\simeq 300$ GeV. Accordingly, somewhat harder spectra are possible in the MAGIC band. 
 
The modeling of the VHE $\gamma$-ray emission from 3C~279 is particularly challenging (Albert et al. 2008a, B\"ottcher et al. 2009).
Indeed, due to the reduced IC scattering efficiency at high energy (e.g. 
Tavecchio \& Ghisellini 2008)
one zone emission models predict a low and rather soft emission in the 
VHE $\gamma$-ray band. 

An additional difficulty for the BLR model  is  the possibility that  VHE photons will be further reduced due to the internal $\gamma \gamma \rightarrow e^{+} \, e^{-}$ absorption with the UV photons of the BLR 
(Liu et al. 2008, Sitarek \& Bednarek 2008, Tavecchio \& Mazin 2008).
In contrast, when the emission occurs outside the BLR, 
 photons with $E<$ 1 TeV are not substantially absorbed (above 1 TeV they will be absorbed interacting with the IR radiation of the torus).
Summarizing, these observations are very difficult to  reconcile with the one-zone framework. Both the EC/BLR and the EC/IR models can reproduce the data but require a rather large flux in the LAT band. Future detections of 3C~279 simultaneously in the LAT  and   VHE band will be crucial to confirm or rule out this possibility.

\subsection{ SED January 2007 }

At the time of the second MAGIC detection, in January 2007, 3C~279 was
in a brighter optical state but in a fainter X-ray state than at the
discovery (February 23, 2006). Motivated by the difficulties of the
one-zone model discussed above (for 2007 the required GeV flux would
be even higher than for 2006), we consider two-zone model and
lepto-hadronic model for this dataset.

\subsubsection{Two-zone Model}

In``two zone'' model, the emission from
the optical up to the X-ray and $\gamma$-ray bands derives from 
a different emission region 
than the VHE $\gamma$-ray photons. As the VHE $\gamma$-ray flare follows the optical one, we assume that the optical up to the X-ray and $\gamma$-ray emission zone is closer to central engine than the VHE $\gamma$-ray emission. Abdo et al. (2010) show that the $\gamma$-ray emission region is located on the parsec scale jet and we therefore model the spectral energy distribution assuming that the optical up to the X-ray and $\gamma$-ray bands are emitted within the BLR while the VHE $\gamma$-ray emission origins    
in a region outside the BLR. This also
minimizes the effect of the KN regime and internal absorption
of $\gamma$-rays. However, the modeling is valid also for emission regions further out in the jet as far as it is still located within IR torus to provide enough seed photons.  

Assuming two regions we are doubling the number of
free parameters, therefore this scenario is far less constrained than
one-zone models. Clearly the magnetic field and particle density will
be lower and the size larger for the more external region. The
parameters for the two regions are reported in Table 4 and the fit to
2007 data in Fig.~9.  The two-zone model can better reproduce the
MAGIC data, due to the possibility to shift the peak of the EC bump to
higher energies.

It is also interesting to speculate about the possible connection
between the energetic flare observed in the IR band and the VHE event,
following the former by almost two days. In the framework of the
two-zone model, an appealing picture assumes that the jet perturbation
(possibly a shell) responsible for the IR flare (assumed to be
produced inside the BLR), travels and eventually reaches the region
where VHE photons can be more easily produced and can escape the
emission region.  From the delay between the two events we can infer
the distance between the location of the infrared to optical
synchrotron and VHE emission zones, $d\simeq c t_{\rm lag} \Gamma^2
\sim2\times 10^{18} $ cm, where we assume that the perturbations
travels with a Lorentz factor $\Gamma\simeq 20$, the value assumed in
the radiative model. Interestingly, this distance is very well
consistent with the assumed size of the VHE $\gamma$-ray emission
zone.

\begin{figure}
\includegraphics[width=0.45\textwidth]{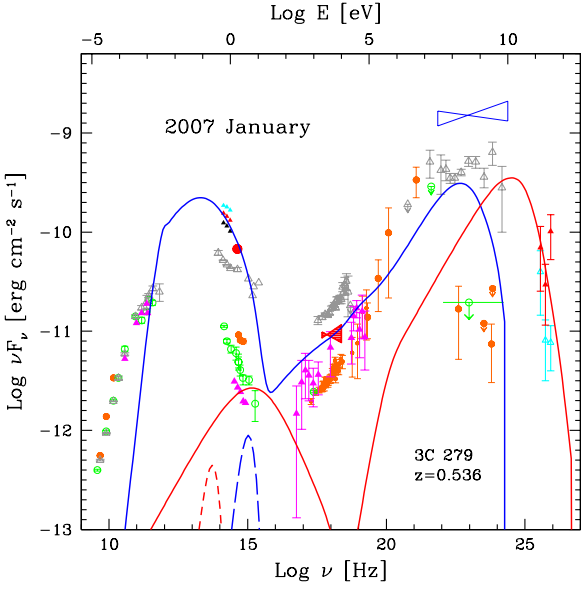}
\caption{The spectral energy distribution of 3C~279 on January 2007 16th (MJD 54116). 
The red symbols show the REM (MJD 54114.4, triangles), KVA (filled circle), RXTE (bow-tie) and 
MAGIC 
(deabsorbed, triangles) data used to fit the 
two-zone model (see text). Blue line shows the emission from the zone inside BLR and red line the emission from outside the BLR.   
Additionally REM data from MJD 54113.3 (Maximum of the REM lightcurve) cyan and MJD 54117.3 (black) are shown. The dashed lines correspond to blackbody radiation from the IR torus (red) and BLR (blue).
The historical data as in Fig.~8.}
\label{sed_2007}
\end{figure}

\subsubsection{Lepto-hadronic model}

Although hadronic models in general are not favored for luminous
quasars (Sikora et al. 2009), they might be viable for 3C~279 as it
does not have as hard X-ray spectrum ($\Gamma<0.5$, for 3C~279
$\Gamma$=0.66 Ballo et al. 2002) as other luminous quasars. Furthermore, 
models taking into account both leptonic and hadronic processes could be viable for describing the spectral energy distribution of quasars.
Therefore, we also fit the data with a lepto-hadronic model. The model will be
described in full detail in a forthcoming paper (R\"uger et al. in prep.).
It comprises a non-thermal proton and
electron distribution injected into the radiation volume as a power law with
lower and upper bound ($\gamma_\textrm{min}$ and $\gamma_\textrm{max}$
respectively). The particles may radiate through a number of channels: The
electrons may radiate through the above described mechanisms of synchrotron
emission and inverse Compton scattering off synchrotron photons. The protons
will also emit synchrotron radiation. Additionally they take part in
photo-hadronic processes. Those are modeled by the cross-sections 
described in Kelner \& Aharonian (2008).
In the photo-hadronic processes $\pi^0$ and $\pi^\pm$ are
produced. The $\pi^0$ decay into $\gamma$-rays may eventually
start a pair cascade in the source if $\gamma$-rays are emitted in the opticallythick regime. The $\pi^\pm$ decays into electrons and positrons which
again emit synchrotron radiation. The time evolution is solved
self-consistently with a scheme described in R\"uger et al. (2010)
for the pure leptonic case.

In this model the number of free parameters is comparable to one-zone
SSC models. We introduce four extra parameters to describe the proton
population. Since we assume identical minimum Lorentz factor and
spectral index for electrons and protons, we end up with a total of
nine free parameters. Unlike the one-zone SSC model the variability
patterns are far more complicated. In this model the lower limit of
variability is still given by the beamed light crossing time, which is
of the order of 4 days. Since electrons and protons may burst
differently and at different times, the determination of the
variability pattern is complicated. If electron and proton density are
increased simultaneously, the electrons produce the well-known soft
lag pattern in the electron synchrotron radiation and the inverse
Compton radiation. In the hadronic component this is also true for the
synchrotron radiation. For the $\pi$ production intermediate secondary
energies peak earlier.

To model the data (fit shown in Fig.~10) we used electron and proton
spectra with a $\gamma_\textrm{min}=150$ and a spectral index
$s=2.2$. For the protons we find $\gamma_\textrm{max}=10^9$, while the
electrons have a much lower $\gamma_\textrm{max}=5\times
10^4$. Spectral breaks are calculated self-consistently from the loss
processes within the radiation zone. The magnetic field is assumed to
be 0.025 G with a source radius of $5.3\times 10^{17}$ cm. The doppler
factor is 42. 
The energy density of the electrons is one order of magnitude higher
than the magnetic field energy density, while the proton energy
density is already six orders of magnitude higher. Even though this is
not in equilibrium this might still be confined, as the proton
gyroradius at highest energies is of the order of the blob radius.

The parameters suggest that the radiation zone may be close to the
end of magnetic confinement zone, $\sim$1000 Schwarzschild radii from
the supermassive black hole. However, the model itself is not limited to that
region, but may be used on every point along the jet axis. 
Although internal pair absorption is included, external photons
producing pairs are not included in the model and therefore if the emission region was inside the BLR the observed flux would be reduced.  

\begin{figure}
\includegraphics[width=0.45\textwidth, angle=270]{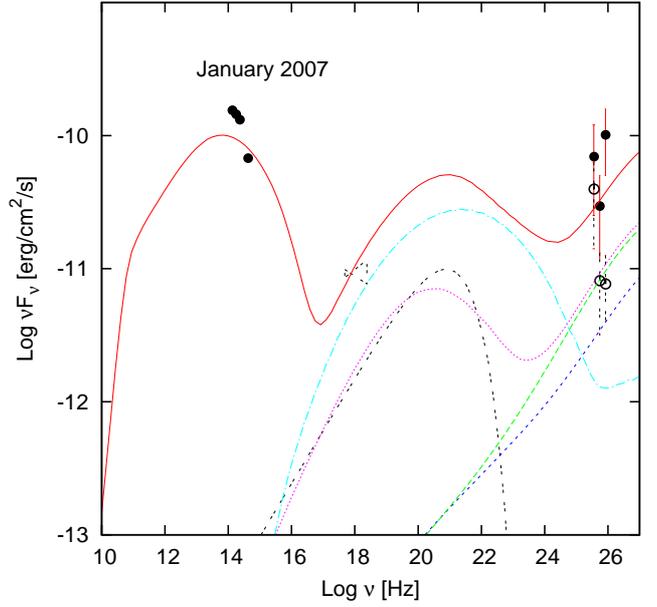}
\caption{The observed spectral energy distribution of January 2007 modeled 
with the lepto-hadronic model. The overall fit is shown with red line with 
the following components at 
high energies: the synchrotron radiation of pair creation electrons and 
positrons cascaded down from the optically thick regime (magenta dotted 
line), synchrotron radiation from positrons from the pion decay (green 
dashed line), synchrotron radiation from electrons from the pion decay (blue dashed line), inverse Compton scattering (cyan dot dashed line) and proton synchrotron emission (black double dashed line). The VHE $\gamma$-ray emission is mostly sum of the three first components while in X-rays the main contribution comes from the inverse Compton scattering (like in the purely leptonic models). The low energy bump is produced by the electron synchrotron radiation.}
\label{spectrum_2007}
\end{figure}

\subsection{ SED January 2009}
For the 2009 observations we present a SED built using
multiwavelength data nearly simultaneous to the MAGIC data taken in the
period 21-31 January, which yielded in an upper limits in the VHE
$\gamma$-ray band. $\gamma$-ray data were derived from {\it Fermi}
LAT and averaged over the same period ({\it Fermi} data were not available
at the previous epochs) and XRT and UVOT data of February 1, 2009 were 
used. The source was in a rather low state at all available
bands. The SED can be modeled quite satisfactorily with a standard one zone
model assuming the emission region inside the BLR (Fig.~11) with typical
parameters (reported in Table 4).

\begin{table*}[th]
\caption{Input parameters for the leptonic emission models. See text for definitions.}
\tiny
\centering
\begin{tabular}{lccccccccccccc}
\hline
\hline
Model       & $\gamma _{\rm min}$ & $\gamma _{\rm b}$& $\gamma _{\rm max}$& $n_1$&$n_2$ &$B$ &$K$ &$R$ & $\delta $ & $\theta$ & $\tau_{\rm BLR}$ &$R_{\rm BLR}$ & $R_{\rm IR}$\\
 & & [$10^3$] & [$10^5$]  &  & &[G] & [$10^4$ cm$^{-3}]$  & $[10^{16}$cm] &  & [deg] & & $[10^{17}$cm] & $[10^{18}$cm]\\\hline
2006 One-zone BLR        							   & 1& 2.5& 3.5& 2& 3.7& 0.15& 2&5 & 20& 2.9 & 0.015& 4& --\\
2006 One-zone IR            							   & 1& 2& 2& 2& 4& 0.19& 1& 4.5& 27& 2 & -- & --& 4\\
\hline
2007 Two-zone: Opt-X-ray zone                            & 1& 0.5& 0.03 & 2& 4.3& 2.2& 5.5& 5& 18&  3.1 & 0.05 & 6& --\\
2007 Two-zone: VHE zone                                     & 45& 20& 5& 2& 4.3& 0.1& 0.01& 10& 18&  3.1 & -- & -- &2.5\\
\hline
2009 One-zone BLR        							                         
& 1& 0.33& 0.2& 2& 3.5& 0.8& 23&3 & 20& 2.1 & 0.1& 6& --\\
\hline
\hline
\end{tabular}
\vskip 0.4 true cm
\label{modelparam}
\end{table*}

\normalsize

\begin{figure}
\includegraphics[width=0.45\textwidth]{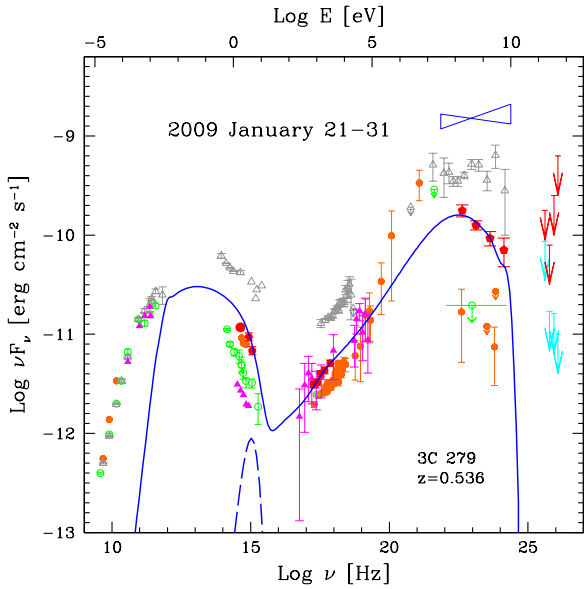}
\caption{The spectral energy distribution of 3C~279 on January 2009: 
MAGIC (taken in the period 21-31 January, red arrows: deabsorbed, cyan arrows: observed), LAT (averaged
over the same period, red pentagons), XRT and UVOT (February 1, 2009 
red squares) and KVA (January 25, 2009 red filled circle). The SED is 
modeled assuming the emission region inside the BLR. The historical data as in Fig.~8. The dashed line correspond to blackbody radiation from the IR torus (red).}
\label{sed_2009}
\end{figure}

\section{Summary and Conclusions}

In this paper we report the MAGIC observations of 3C~279 in January
2007 and from December 2008 to April 2009. The January 2007
observations yielded a detection on the night of January 16,
when the source was in a high optical state. Like in February 2006
when the source was discovered, in January 2007 the source was
detected on one night only. However, in January 2007 MAGIC observed
the source also in the previous and the following night and these
observations resulted in non-detection, which confirms the day-scale
variability of the source.

In February 2006 and January 2007 the source was in a high optical-IR
state and was also showing a rather high state in X-rays. However, no
major increase in the optical or X-ray flux was seen simultaneously to 
the VHE $\gamma$-ray flare. Unlike in February 2006 and January 2007, 
in January 2009 the source was observed by MAGIC during a low optical-X-ray
state. The existence of LAT data for this epoch shows that the source
was relatively faint also at GeV energies. Consistently, the MAGIC
observations resulted in upper limits.

In the spring of 2009 {\it Fermi} detected a fast $\gamma$-ray flare
from 3C~279 which was coincident with a rotation of the optical
polarization angle of 180 degrees that had duration of $\sim 20$
days. In early 2007 such rotation of the polarization angle also took
place (Larionov et al. 2008), but with rotation of $\sim 360$ degrees,
the duration of rotation $\sim$2 times longer and the direction of the
rotation opposite to what was seen in 2009. The MAGIC detection in
January 2007 took place in the beginning of the rotation of optical
polarization angle, thus confirming that rotation of optical
polarization angle in 3C~279 are recurrently accompanied with
$\gamma$-ray flares. This seems to be common in FSRQs, the same
behavior has been observed in PKS~1510-089 in 2009 (Marscher et
al. 2010b). 

However, the patterns in 3C~279 in January 2007 and February 2009 are
clearly different, which might suggest that different mechanisms for
producing the rotation of the polarization angle could be in action
(e.g. for 2009 the rotation could be produced by turbulence, see
e.g. D'Arcangelo et al. 2007).Therefore it is not possible to draw
definite conclusions on the connection between the rotation of
polarization angle with $\gamma$-ray flares either.

We find that the spectral energy distribution of 3C~279 in 2007
January cannot be explained by the standard one-zone SSC+EC model,
where the emission region is located inside the BLR which provides the
seed photons for inverse Compton scattering. Instead we explore a
two-zone model, where the VHE $\gamma$-ray emitting region is located
just outside the BLR, while the standard optical-to-X-ray and
$\gamma$-ray emitting region is still inside the BLR region and a
lepto-hadronic model, which both fit the data reasonably well. Also
other possible scenarios for the emission exists. Sitarek \& Bednarek
(2010) suggested a leptonic cascade model. Several authors have also
suggested that blazar emission zone would be located at parsec scale
distances from the black hole: there $\gamma$-rays could be produced
via up-scattering of infrared photons from the hot dust (Blazejowski
et al. 2000, Sikora et al. 2009). It has also been suggested that the
relativistic jet could drag part of the BLR clouds to large distances
from the central engine (at least in radio galaxies: Arshakian et
al. 2010, Leon-Tavares et al. 2010a) and in principle these photons
could serve as external seed photons for inverse Compton
scattering (Leon-Tavares et al. 2011). The multiwavelength behavior of the source in January 2007
can also be interpreted in favor of emission zone beyond VLBI core
and in such case multi emission shell model suggested by Marscher et
al. (2010) could be viable.

After the first detection of 3C~279 by MAGIC, two other flat spectrum
radio quasars have been detected in the VHE $\gamma$-ray band, PKS~1510-089
($z=0.36$, Wagner et al. 2010)
and PKS~1222+21 ($z=0.432$,
Mariotti et al. 2010, Aleksi\'c et al. 2011). In all the cases the emission of the VHE $\gamma$-ray
radiation poses problems to the standard models, as discussed
in this paper and in Aleksi\'c et al. (2011). Future multiwavelength observations, including higher sensitivity observations with MAGIC II, will be needed to
improve our understanding of the high-energy emission from FSRQs.

\begin{acknowledgements}
We would like to thank the Instituto de Astrof\'{\i}sica de
Canarias for the excellent working conditions at the
Observatorio del Roque de los Muchachos in La Palma.
The support of the German BMBF and MPG, the Italian INFN, 
the Swiss National Fund SNF, and the Spanish MICINN is 
gratefully acknowledged. This work was also supported by 
the Marie Curie program, by the CPAN CSD2007-00042 and MultiDark
CSD2009-00064 projects of the Spanish Consolider-Ingenio 2010
programme, by grant DO02-353 of the Bulgarian NSF, by grant 127740 of 
the Academy of Finland, by the YIP of the Helmholtz Gemeinschaft, 
by the DFG Cluster of Excellence ``Origin and Structure of the 
Universe'', and by the Polish MNiSzW Grant N N203 390834. Authors wish to thank Alan Marscher for useful discussions and the referee Elena Pian for useful comments on the earlier version of the paper.
\end{acknowledgements}


\begin{thebibliography}{}


\bibitem[2009]{brightsource}
Abdo, A.\ et al.\ (the Fermi LAT Collaboration) 2009, ApJ, 700, 597

\bibitem[2009]{Nature}
Abdo, A. \ et al.\ (the Fermi LAT Collaboration and Members of the 3C~279 Multi-Band Campaign), 2010, Nature, 463, 919 

\bibitem[2011]{agudo}
Agudo, I., Jorstad, S., Marscher, A. et al. 2011, ApJL 726, 13

\bibitem[2009]{pks2155}
Aharonian, F.\ et al.\ (the HESS Collaboration) 2009, A\&A, 502, 749

\bibitem[2007]{mrk501}
Albert, J.\ et al.\ (the MAGIC Collaboration) 2007, ApJ, 669, 862

\bibitem[2008a]{science}
Albert, J.\ et al.\ (the MAGIC Collaboration) 2008a, Science, 320, 1752

\bibitem[2008b]{crab}
Albert, J.\ et al.\ (the MAGIC Collaboration) 2008b, ApJ, 674, 1037

\bibitem[2008c]{NIMA}
Albert, J.\ et al.\ (the MAGIC Collaboration) 2008c, Nucl. Instr. Meth. 
A, 594, 407 

\bibitem[2008d]{randomforest}
Albert, J.\ et al.\ (the MAGIC Collaboration), 2008d, Nucl. Instr. Meth. A, 588, 424

\bibitem[2010]{mrk421}
Aleksi\'c et al. (the MAGIC Collaboration), 2010, A\&A, 519, 32

\bibitem[2011]{1222}
Aleksi\'c et al. (the MAGIC Collaboration), 2011, ApJL, 730, 8

\bibitem[2009]{timing}
Aliu, E. \ et al.\ (the MAGIC Collaboration), 2009, Astropart. Phys., 30, 293

\bibitem[2010]{arshakian}
Arshakian, T. G., Leon-Tavares, J., Lobanov, A. P. et al. 2010, MNRAS, 401, 1231

\bibitem[2002]{ballo}
Ballo, L., Maraschi, L., Tavecchio, F. et al. 2002, ApJ, 567, 50

\bibitem[1979]{bessell}
Bessell, M. S. 1979, PASP, 91, 589

\bibitem[2000]{blazejowski}
Blazejowski, M., Sikora, M., Moderski, R., Madejski, G. M. 2000, ApJ, 545, 107	

\bibitem[2009]{boettcher} 
B\"{o}ttcher, M., Reimer, A., Marscher, A. 2009, ApJ, 703, 1168

\bibitem[2009]{britzger}
Britgzer, D., Carmona, E. Majumdar, P. et al., Proc. 31st ICRC, 
Lodz, Poland, July 2009, arXiv:0907.0973

\bibitem[2003]{bretz03}
Bretz, T. \& Wagner, R., 2003, in Proceedings of the 28th International Cosmic
   Ray Conference, Tsukuba, Japan, 5, 2947

\bibitem[2008]{bretz08}
Bretz, T. \& Dorner, D. (MAGIC Collab.), 2008, AIP Conf. Proc. 1085, 664

\bibitem[2008]{chatterjee2008} 
Chatterjee R., Jorstad , S. G., Marscher, A. 2008, ApJ. 689, 79

\bibitem[2004]{chary}
Chary R., Casetarno, S., Dickinson, M. E. 2004, ApJS, 154, 80

\bibitem[2003]{Chi03} 
Chincarini, G., Zerbi, F. M., Antonelli, L. A., et~al.\ 2003, The Messenger, 113, 40

\bibitem[{Ciprini et~al.} {2008}]{ATel}
Ciprini, S. et al. (Fermi Collaboration), 2008, ATel 1864

\bibitem[2004]{collmar}
Collmar, W. et al. 2004, Proceedings of the 5th INTEGRAL Workshop on the INTEGRAL 
Universe (ESA SP-552). 16-20 February 2004, Munich, Germany. Scientific 
Editors: V. Schönfelder, G. Licht \& C. Winkler, p.555

\bibitem[2010]{collmar}
Collmar, W., B\"ottcher, M., Krichbaum, T. et al. 2010, A\&A, 522, 66

\bibitem[2004a]{Cov04a} 
Covino, S., Zerbi, F. M., Chincarini, G., et~al.\ 2004a, AN, 325, 543

\bibitem[2004b]{Cov04b} 
Covino, S., Stefanon, M., Sciuto, G., et~al.\ 2004b, SPIE, 5492, 1613

\bibitem[1997]{daum} 
Daum, A. et~al. (The HEGRA Collaboration), 1997, APh, 8, 1

\bibitem[2007]{darcangelo}
D'Arcangelo, F., Marscher, A. P., Jorstad, S. et al. 2007, ApJ, 659, 107

\bibitem[2007]{degl}
Degl'Innocenti, E. L., Bagnulo, S., Fossati, L. 2007, in: The Future of Photometric, Spectrophotometric, and Polarimetric Standardization, editor: C. Sterken, ASP 364, 495

\bibitem[1993]{dermer93} 
Dermer, C. D. \& Schlickeiser, R. 1993, ApJ, 416, 458 

\bibitem[1997]{Dev97} 
Devillard, N. 1997, The Messenger, 87

\bibitem[2010]{dominguez}
Dom\'inguez, A., Primack, J. R., Rosario, D. J., et al. 2010, MNRAS, 410, 2556

\bibitem[2004]{fazio}
Fazio G.~G., et al., 2004, ApJS, 154, 39

\bibitem[1994]{fomin}
Fomin, V. P. et al., 1994, Astropart. Phys., 2, 137



\bibitem[2004]{2004ApJ...611.1005G}  
Gehrels, N., et al.\ 2004, \apj, 611, 1005 

\bibitem[2008]{ghisellini08} 
Ghisellini, G. \& Tavecchio, F. 2008, MNRAS, 387, 1669 

\bibitem[2009]{ghisellini09}
Ghisellini, G. \& Tavecchio, F. 2009, MNRAS, 397, 985

\bibitem[1992]{hartman1992} 
Hartman, R. C. et al. 1992, ApJ 385L, 1H     
  
\bibitem[1996]{hartman1996}
Hartman, R. C. et al.\ 1996, ApJ, 461, 698 

\bibitem[2001]{hartman2001}
Hartman, R. C. et al. 2001, ApJ 558, 583H

\bibitem[1985]{hillas}
Hillas, A. M., 1985, in Proceedings of 
the 19th International Cosmic Ray Conference, La Jolla, 3, 445

\bibitem[1993]{hewittburbridge1993}
Hewitt, A. \& Burbidge G. 1993, ApJS, 87, 451

\bibitem[2001]{jorstad2001}
Jorstad, S. G., Marscher, A. G., Mattox J. R. et al. 2001, ApJ, 556, 738
	
\bibitem[2004]{Jorstad2004}
Jorstad, S. G. et al. 2004, AJ. 127, 3115

\bibitem[2010]{jorstad2010}
Jorstad, S. G., Marscher, A. et al. 2010, 2009 Fermi Symposium and eConf Proceedings, arXiv:0912.5230 

\bibitem[2005]{Kalberla}
Kalberla, P. M. W. et al. 2005, A\&A, 440, 775 

\bibitem[2008]{2008PhRvD..78c4013K}
Kelner, S.~R. \& Aharonian, F.~A. 2008, Phys. Rev. D. 78, 4013

\bibitem[2010]{leon}
Leon-Tavares, J., Lobanov, A. P., Chavushyan, V. H. et al. 2010, ApJ, 715, 355

\bibitem[2011]{leon01}
Leon-Tavares, J., Valtaoja, E., Tornikoski, M., L\"ahteenm\"aki, A. \& Nieppola, E. 2011, submitted to A\&A, arXiv:1102.1290v1

\bibitem[2001]{lessard}
Lessard, R. W. et al., 2001, Astropart. Phys., 15, 1

\bibitem[1983]{lima} 
Li,T.-P., and Ma, Y.-Q., 1983
ApJ,272, 317.

\bibitem[2008]{liu}
Liu, H. T., Bai, J. M. \& Ma, L. 2008, ApJ, 688, 148

\bibitem[2006]{Lindfors2006}
Lindfors, E., T\"urler, M., Valtaoja, E. et al. 2006, A\&A 456, 895

\bibitem[1992]{mannheim} 
Mannheim, K., \& Biermann, P.~L. A\&A, 1992, 253, L21

\bibitem[1992]{maraschi92} 
Maraschi, L., Ghisellini, G., \& Celotti, A.\ 1992, ApJL, 397, L5

\bibitem[1994]{maraschi94}
Maraschi, L. et al. 1994, ApJ, 435, 91

\bibitem[2003]{maraschi03}
Maraschi, L. \& Tavecchio, F. 2003, ApJ, 593, 667

\bibitem[2010]{marscher}
Marscher, A. P. \& Jorstad, S. G. 2010a, in the Proceedings of Fermi meets Jansky – AGN in Radio and Gamma-Rays editors:
Savolainen, T., Ros, E., Porcas, R.W., \& Zensus, J.A.

\bibitem[2010]{marscher}
Marscher, A. P., Jorstad, S. G., Larionov, V. et al. 2010b, ApJL, 710, L126

\bibitem[2010]{pks1222}
Mariotti, M. (for the MAGIC Collaboration) 2010, ATel 2684

\bibitem[]{mattox}
Mattox, J. R., Bertsch, D. L., Chiang, J., et al. 1996, ApJ, 461, 396

\bibitem[2003]{Muecke2003}
M\"ucke, A.\ et al. 2003, APh, 18, 593

\bibitem[2008]{nenkova}
Nenkova, M., Sirocky, M., Ivezic, Z. \& Elitzur, M. 2008, ApJ, 685, 147

\bibitem[1992]{1992ApJ...395..130P}
Pei, Y.~C.\ 1992, ApJ, 395, 130 

\bibitem[1999]{pian}
Pian, E., Urry, C. M., Maraschi, L. et al. 1999, ApJ, 521, 112

\bibitem[2008]{2008MNRAS.383..627P} 
Poole, T.~S., et al.\ 2008, MNRAS, 383, 627 

\bibitem[2005]{primack}
Primack, J. R., Bullock, J. S. Somerville, R. S., 2005, in High Energy Gamma-Ray Astronomy American Institute of Physics Conference Series, editors: F. Aharonian, H. Voelk, D. Horns (AIP, Heidelberg), 745, 23


\bibitem[1998]{raiteri}
Raiteri, C. M., Villata, M., Lanteri, L., Cavallone, M. \& Sobrito, G. 1998, A\&AS 130, 495

\bibitem[2005]{riegel}
Riegel, B., et al. (MAGIC Collab.), 2005, in Proceedings of the 29th
International Cosmic Ray Conference, Pune, India, 5, 215

\bibitem[2010a]{2010MNRAS.401..973R}
R{\"u}ger, M., Spanier, F. \& Mannheim, K. 2010, MNRAS, 401, 973


\bibitem[1998]{1998ApJ...500..525S}
Schlegel, D. J., Finkbeiner, D. P.\& Davis, M., 1998, ApJ, 500, 525

\bibitem[1994]{sikora94} 
Sikora, M., Begelman, M.~C., \& Rees, M.~J.\ 1994, \apj, 421, 153 

\bibitem[2001]{Sikora2001}     
Sikora, M. et al. 2001, ESASP 459, 259S.

\bibitem[2008]{sikora08} 
Sikora, M., Moderski, R., \& Madejski, G.~M.\ 2008, \apj, 675, 71

\bibitem[2009]{sikora09}
Sikora, M., Stawarz, L. Moderski, R. Nalewajko, K. \& Madejski, G. 2009, 
ApJ, 704, 38

\bibitem[2008]{sitarek}
Sitarek, J. \& Bednarek, W. 2008, MNRAS, 391, 624

\bibitem[2010]{sitarek10}
Sitarek, J. \& Bednarek, W. 2010, MNRAS, 409, 662

\bibitem[2008]{tavecchio08} 
Tavecchio, F.\& Ghisellini, G. 2008, MNRAS, 386, 28

\bibitem[2009]{tavecchio09} 
Tavecchio, F., Ghisellini, G. 2009, MNRAS, 394, 131

\bibitem[2009]{daniel}         
Tavecchio, F. and Mazin, D. 2009, MNRAS 392, 40

\bibitem[1990]{turnsheak}
Turnsheak, D. A., Bohlin, R. C. Williamson, R. L. et al. 1990, AJ, 99, 1243


\bibitem[2010]{pks1510}
Wagner, S. J. (the HESS collaboration) 2010, HEAD Meeting, Hawaii (BAAS, 42, 2, 07.05)

\bibitem[1998]{wehrle}
Wehrle, A. et al. 1998, ApJ, 497, 178

\bibitem[2001]{Zer01} 
Zerbi, F. M., Chincarini, G., Ghisellini, G., et~al.\ 2001, AN, 322, 275

\end{thebibliography}
\end{document}